\documentclass[aip,jcp,a4paper,amsmath,amssymb,twocolumn,10pt]{revtex4-1}
\usepackage{graphicx}
\usepackage{dcolumn}
\usepackage{bm}
\usepackage{color}

\DeclareMathOperator{\erfc}{erfc}

\begin{document}

\title{\textcolor{blue}{Simulations of Nanocrystals Under Pressure: Combining
Electronic Enthalpy and Linear-Scaling Density-Functional Theory}}

\author{Niccol\`{o} R. C. Corsini}

\email{niccolo.corsini@imperial.ac.uk}

\affiliation{Department of Physics and Department of Materials, Imperial College
London, Exhibition Road, London SW7 2AZ, UK.}

\author{Andrea Greco}

\affiliation{Department of Physics and Department of Materials, Imperial College
London, Exhibition Road, London SW7 2AZ, UK.}

\author{Nicholas D. M. Hine}

\affiliation{Department of Physics and Department of Materials, Imperial College
London, Exhibition Road, London SW7 2AZ, UK.}

\affiliation{Cavendish Laboratory, J. J. Thompson Avenue, Cambridge CB3 0HE, UK.}

\author{Carla Molteni}

\affiliation{King's College London, Department of Physics, Strand, London WC2R
2LS, UK.}

\author{Peter D. Haynes}

\affiliation{Department of Physics and Department of Materials, Imperial College
London, Exhibition Road, London SW7 2AZ, UK.}

\date{\today}
\begin{abstract}
We present an implementation in a linear-scaling density-functional
theory code of an electronic enthalpy method, which has been found
to be natural and efficient for the \emph{ab initio} calculation of
finite systems under hydrostatic pressure. Based on 
a definition
of the system volume as that enclosed within an electronic density
isosurface {[}Phys. Rev. Lett., \textbf{94}, 145501 (2005){]}, it
supports both geometry optimizations and molecular dynamics simulations.
We introduce an approach for calibrating the parameters defining the
volume in the context of geometry optimizations and discuss their
significance. Results in good agreement with simulations using explicit
solvents are obtained, validating our approach. Size-dependent pressure-induced
structural transformations and variations in the energy gap of hydrogenated
silicon nanocrystals are investigated, including one comparable in
size to recent experiments. A detailed analysis of the polyamorphic
transformations reveals three types of amorphous structures and their
persistence on depressurization is assessed.
\end{abstract}
\maketitle

\section{Introduction}

In recent years, the study of nanomaterials under pressure has acquired
increased scientific and technological importance.\cite{San-Miguel2006a}
In part due to their 
large ratio of surface to volume atoms, nanocrystals
display a host of properties that differ from those of their bulk
counterparts.\cite{Tolbert1995a} New dimensions are added to phase
diagrams when the sizes, surface reconstructions and terminations
of the nanocrystals are taken into account.\cite{Tolbert1996a} This
has generated particular interest in nanocrystals displaying quantum
confinement with diverse applications ranging from biomarkers to quantum
transistors.\cite{alivisatos1996perspectives,alivisatos2005quantum,loss1998quantum}
There is great technological potential in the possibility of using
pressure to tune the
physical properties of semiconducting nanocrystals, that can depend
sensitively on their structures.\cite{alivisatos1996perspectives}
Attaining such control at the nanoscale holds the promise of 
novel technological applications such as tunable photovoltaic devices,
shock-absorbers,\cite{Cococcioni2005} and nanoscale stress sensors\cite{choi2009strain,choi2010luminescent}.
The additional surface effects also open the door to transformation
pathways that are not available to the bulk material, potentially
allowing a system to become trapped in metastable states with novel
properties.\cite{chen1997size,herhold1999structural,grunwald2012metastability}
Moreover, sufficiently small nanocrystals can be synthesized with few or
no defects and are thus ideal models to study the kinetics of solid-solid
phase transitions.\cite{Jacobs2001,Jacobs2001a} Recently, progress
has been made in directly observing structural transformations in
nanocrystals\cite{zheng2011observation} and bulk single crystals\cite{Shen2012}.
Direct monitoring of transformation pathways in nanosystems is, however,
still challenging with the resolution of existing experimental probes
and understanding can thus greatly benefit from the insights that computer
simulations provide. 
The nanocrystals of interest here are of an intermediate
size: larger than molecules but too small to be treated satisfactorily
with macroscopic concepts such as strain and stress fields in continuum
models. An atomistic treatment is crucial to capture the details of the structural changes, including
the shape and surface effects.

While empirical potentials are good for modelling a variety of materials,
the complex bonding rearrangements associated with structural transformations
of materials such as covalent semiconductors mean that \emph{ab initio}
methods such as density-functional theory (DFT) are essential to capture
the details of the structure and dynamics with accuracy. However,
the large length- and time-scales associated with the structural transformations
of experimentally relevant systems pose a significant computational
challenge. The $\mathcal{O}(N^{3})$ scaling of the computational
effort in traditional methods such as the plane-wave pseudopotential
(PWPP) formulation of DFT limits the number of atoms $N$ that can
be simulated to a few hundred and thereby seriously constrains the
attainable sizes of nanocrystals. This can be addressed by working
with a linear-scaling DFT code such as ONETEP,\cite{Skylaris2005a}
for which the favorable balance of cost and accuracy allows the investigation
of nanocrystals with many thousands of atoms.\cite{Hine2009,Avraam2012a}
Even then, the challenge persists of modelling the pressure transmission
between solvent molecules and nanocrystals---in analogy to experiments
where nanocrystals are dissolved and placed under pressure in a diamond
anvil cell. The many degrees of freedom comprising realistic solvents
and the many solvent-nanocrystal collisions that need to be averaged
over to sample the appropriate thermodynamic ensemble exclude a full
\emph{ab initio} treatment. One approach to tackle this challenge
is to retain an explicit description of the solvent by embedding an
\emph{ab initio} simulation of the nanocrystals within a cheaper classical
description of the solvent.\cite{1martonak2000ab,1martovnak2001new,1molteni2001first,1Martonak2002,1molteni2005polyamorphism,morgannanoletter2004,morgan2007pressure}
However, sampling rare events such as structural transformations happening
over long time-scales, whilst retaining a sufficiently short time
step to describe the solvent-nanocrystal collisions, generally requires
unfeasibly large numbers of molecular dynamics (MD) 
steps to be performed.
Transformations can be obtained within shorter simulation times by
over-pressurizing the systems but comparability with experiment is
hindered in the process. Approaches exist to surmount this issue by
accelerating the free energy landscape exploration and have been applied
to the pressure-induced structural transformations of nanocrystals\cite{grunwald2007efficient,grunwaldnanoletter2009}
and bulk crystals.\cite{martonakreview2005,martonakreview2011,1bealing2009pressure,1bealing2010wurtzite}
In practice, however, these remain computationally demanding.

Alternatively, constant pressure simulations of finite systems can
be performed, in both MD and quasistatic geometry optimization, by
directly optimizing the enthalpy once a suitable definition for the
finite volume has been made. This can be done in a variety of ways
in terms of atomic or electronic coordinates leading to an implicit
description of the solvent. Some examples of total volume definitions
have been suggested in terms of: a sum of atomic volumes,\cite{sun2002new}
a function of the average inter-particle distance,\cite{landau2002new}
the inertia tensor eigenvalues\cite{sun1998structural,1bealing2010constant}
and the smallest convex polyhedron to circumscribe all surface atoms.\cite{calvo2004pressure}
These different approaches have been compared elsewhere
and were shown to qualitatively reproduce results obtained with explicit
solvents.\cite{1baltazar2006assessment} By working with quasistatic
geometry optimizations at zero temperature, one removes the need for
equilibration with barostats and thermostats thereby giving a comparatively
inexpensive way of sampling the enthalpy landscape. Depending on system complexity, 
this may not
give a globally optimized structure nor precise information on transition
paths; however, it provides the structure and energetics of the nearest
local minimum.

In the present work we use an electronic enthalpy functional $H=U+P_{{\rm in}}V_{{\rm e}}$,
where $U$ is the total Kohn-Sham internal energy of the system, $P_{{\rm in}}$
the input pressure and $V_{{\rm e}}$ a volume definition based on
an electronic-density isosurface.\cite{Cococcioni2005} The latter
allows for the description of complex geometries and the enthalpy
is optimized within the linear-scaling DFT code ONETEP. We introduce
an approach to calibrate the parameters defining $V_{{\rm e}}$ in
the context of geometry optimizations and use it to simulate pressure-induced
structural transformations in hydrogenated Si nanocrystals. Our results
are comparable to those obtained with other methods\cite{Cococcioni2005,1molteni2001first,1molteni2005polyamorphism}
and validate our approach. Si nanocrystals are of intrinsic interest
due to their potential to overcome the indirect character of the lowest-energy
interband transition and to be useful in optoelectronic devices.\cite{anthony2009photoluminescence,jurbergs2006silicon,wilson1993quantum}
Recently, Si nanocrystals with structures based on high-pressure bulk
phases have been proposed as candidates for photovoltaic applications
as they display multi-exciton generation and high quantum efficiencies.\cite{wippermann2013high}
They are also found to transform under pressure between a variety
of crystalline and amorphous structures that are still the subject
of theoretical and experimental investigations in both the porous
and collolloidal forms. ${\rm Si}_{181}{\rm H}_{110}$, the largest
nanocrystal in the present work, of diameter 2.2~nm, is comparable
in size to experimentally-tested organically passivated colloidal
nanocrystals\cite{Hannah2012a} and demonstrates the capability of
our approach. 

\section{Methodology}

The linear-scaling DFT code ONETEP is based on the single particle
density-matrix (DM) $n({\bf r},{\bf r'})$ formulation of the Kohn-Sham
equations in terms of a set of local orbitals $\left\{ \phi_{\alpha}({\bf r})\right\} $,
referred to as non-orthogonal generalized Wannier functions (NGWFs).
These are
spatially localized within spheres of radii $\left\{ R_{\alpha}\right\} $
centered on the atomic coordinates as
\begin{equation}
n({\bf r},{\bf r'})=\sum_{\alpha\beta}\phi_{\alpha}({\bf r})K^{\alpha\beta}\phi_{\beta}^{*}({\bf r'}),
\end{equation}
where $K^{\alpha\beta}$ is called the density-kernel. The electronic
density $\rho({\bf r})$ is related to the DM by $\rho({\bf r})=2n({\bf
  r},{\bf r})$, where the factor of two accounts for the spin
degeneracy. The NGWFs are themselves expanded in terms of a fixed
underlying basis of psinc functions equivalent to a systematic
plane-wave basis.\cite{Mostofi2002} In the course of a calculation the
total energy is minimized with respect to both $K^{\alpha\beta}$ and
$\left\{ \phi_{\alpha}\right\} $ in two nested loops, subject to the
constraints of normalization and idempotency.\cite{mcweeny1960} Linear
scaling is achieved by exploiting the property of nearsightedness that
allows the DM to be truncated for systems with an energy
gap.\cite{prodan2005} The electronic structure can then be described
with plane-wave accuracy in terms of a minimal basis of \emph{in situ}
optimized NGWFs.

It has been shown that the elastic properties of bulk Si in the
diamond phase calculated with ONETEP and the PWPP code
CASTEP\cite{clark2005} give equivalent
results\cite{skylaris2007achieving} when using the same
norm-conserving Si pseudopotential\cite{lin1993}, local density approximation
exchange-correlation functional\cite{ceperley1980ground,perdew1981self} and plane-wave cutoff $E_{{\rm c}}$.
Beyond the fact that it is well-described by DFT, Si was used as a
test system here due to the plethora of experimental data and
computational studies with different pressure methods available for
comparison.  

For the calibration we require a reference DFT bulk
modulus at zero pressure $B_{0}$ and its pressure derivative
$B_{0}'=\partial B/\partial P|_{P=0}$. This was calculated with CASTEP
using a 2-atom primitive simulation cell, a grid of $8\times8\times8$
k-points and $E_{{\rm c}}=800$~eV. By fitting the universal Vinet
equation of state\cite{Vinet1986,vinet1999universal} we obtained
values of $B_{0}=96.85$~GPa and $B_{0}'=4.08$.

The local orbital approach has the advantage that the $\left\{ \phi_{\alpha}\right\} $
are strictly zero on all grid points outside the localization radii\cite{note1} and vacuum comes at a negligible computational overhead. This is particularly advantageous for finite systems as interaction with periodic
images can easily be eliminated.\cite{hine2011electrostatic} Pulay
corrections to the Hellmann-Feynman forces are required to achieve
accurate ionic forces and optimized structures.\cite{ruiz2012pulay,hine2011accurate}
The use of \emph{in situ} optimized orbitals reduces the egg-box
effect\cite{soler2002siesta} observed in fixed orbital approaches. 

The nanocrystals were quasistatically relaxed at different pressures
using the quasi-Newton BFGS algorithm for geometry optimization.\cite{nocedal1999numerical}
The parameters which control the accuracy of the geometry optimization
must be carefully chosen for the calculations to be converged and
the structures correctly relaxed. Unless specified otherwise we used
$E_{{\rm c}}=800$~eV, a universal NGWF radius $R_{\phi}=8\,a_0$,
an atomic displacement tolerance of $10^{-2}\,a_0$, an energy
tolerance per atom of $2\times10^{-5}$~Ha and a force tolerance
of $10^{-3}\,{\rm Ha}~a_0^{-1}$. 

\begin{figure}
\includegraphics[scale=0.425]{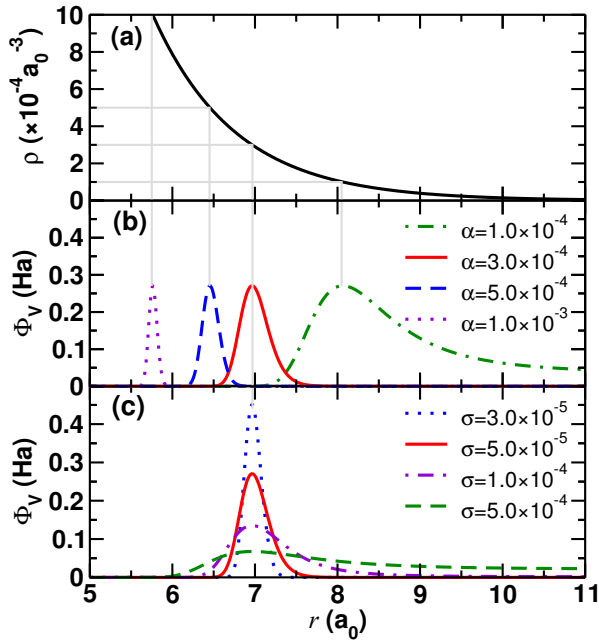}
\caption{\label{fig:Potentials-1}(a) Normalized hydrogenic 1$s$
  electronic density $\rho(r)$; (b) resulting potential
  $\Phi_V(r)$ for different values of $\alpha$ with constant
  $\sigma=5\times10^{-5}~a_0^{-3}$ and pressure $P_{\rm in}=1$~GPa; (c)
  $\Phi_{V}(r)$ at constant $\alpha=3\times10^{-4}~a_0^{-3}$ for
  different values of $\sigma$ again with $P_{\rm in}=1$~GPa.}
\end{figure}

The electronic enthalpy method to simulate finite systems under external
pressure proposed by Cococcioni \emph{et~al.}\cite{Cococcioni2005}
introduces a thermodynamic functional $H=U+P_{\rm in}V_{\rm e}$
which can be minimized self-consistently within DFT algorithms. $V_{\rm e}$
is defined as the interior of an electronic-density isosurface at
a chosen cutoff density $\alpha$. Introducing the Heaviside
step function in terms of density values $\theta(\rho)$, $V_{\rm e}$
is calculated as
\begin{equation}\label{eq:Vdef}
V_{\rm e}=\int\theta\left(\rho({\bf r})-\alpha\right)\,{\rm {\rm d}}{}^{3}r.
\end{equation}
For computational purposes, the step function can be approximated
by the complementary error function as
\begin{equation}
\theta(\rho({\bf r})-\alpha)\simeq\frac{1}{2}\erfc\Biggl(\frac{\alpha-\rho({\bf r})}{\sigma\sqrt{2}}\Biggl).
\end{equation}
The parameter $\sigma$ adjusts the sharpness of the step function
and plays an important role for numerical reasons. The resulting potential
contribution is
\begin{align} \label{eq:phiv}
\Phi_{V}({\bf r}) & =P_{\rm in}\,\frac{\delta V_{\rm e}}{\delta\rho({\bf r})}\nonumber \\
&= \frac{P_{\rm in}}{\sigma\sqrt{2\pi}}\exp\left(-\frac{(\rho({\bf r})-\alpha)^{2}}{2\sigma^{2}}\right).
\end{align}
Since the potential does not explicitly depend on the nuclear positions,
the compression is implicitly transmitted to the nuclei by virtue
of the forces obtained from the Hellmann-Feynman theorem\cite{feynman1939forces}.
This can be related to the effect of $\Phi_{V}$, which for a decaying
density profile as in Fig.~\ref{fig:Potentials-1}, is a distorted
Gaussian in real space and favours the compression of the electronic
density for positive pressures. The shape of $\Phi_{V}$ is determined
by the pair of input parameters $\alpha$ and $\sigma$, and approximates
the solvent-nanocrystal interaction. $\alpha$ defines the excluded
volume of the solvent molecules and $\sigma$ controls the range and
intensity of interaction in a 
manner reminiscent of soft-sphere potentials.
While the method describes the solvent implicitly, providing a homogeneous
and time-averaged description, the emphasis is laid on the role played
by electrons as pressure mediators with an account of the shape of
the nanocrystal as the pressure 
is applied normal to the isosurfaces.  This results in a natural
description that allows the seamless modelling of the excluded volume
of intricate nanocrystal geometries.  
It also removes the need for 
equilibration with barostats
and focuses the computational effort on the electronic structure of
the nanocrystal.

\begin{figure}
  \includegraphics[scale=0.45]{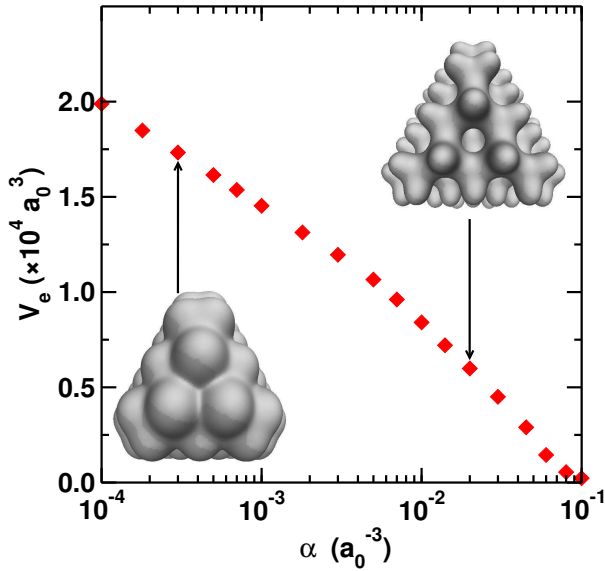}\caption{\label{fig:Vol-1-1}Electronic
    volume $V_{\rm e}$ as a function of the density cutoff $\alpha$
    for ${\rm Si}_{71}{\rm H}_{60}$ relaxed at 0~GPa and the
    corresponding isosurfaces for $\alpha=3\times10^{-4}~a_{0}^{-3}$
    and $\alpha=2\times10^{-2}~a_{0}^{-3}$.}
\end{figure}

Figure~\ref{fig:Vol-1-1} shows the isosurface bounding $V_{\rm e}$
for a range of values of $\alpha$. For larger values of $\alpha$, the isosurface
describes voids inside the nanocrystal, revealed by changes in slope
in the plot, and results in pressure being induced internally which
compensates the applied pressure. In order to describe a realistic
solvent, $\alpha$ has to be chosen sufficiently small to apply a
homogeneous compression without describing the rugosities of the nanocrystal
too closely. When going to very small values of $\alpha$, unphysically
large excluded volumes are obtained. Figure~\ref{fig:Potentials-1}
also shows that for a given choice of $\alpha=3\times10^{-4}~a_0^{-3}$,
$\sigma=5\times10^{-4}~a_0^{-3}$ leads to a $\Phi_{V}({\bf r})$
that clearly fails to vanish at large radii (Fig.~\ref{fig:Potentials-1}c).
From Eq.~\ref{eq:phiv} it is evident that far from the nanocrystal
where $\rho \rightarrow 0$, the potential $\Phi_V \sim (P_{\rm
  in}/\sigma) \exp(-\alpha^2/2\sigma^2)$, and therefore a sufficiently
small value of $\sigma/\alpha$ must therefore be chosen for the excluded
volume to be well-defined. However, a sufficiently large value of
$\sigma$ must be chosen for the potential $\Phi_{V}({\bf r})$ to
be accurately integrated on the underlying real-space grid which has
a spacing of $\Delta=0.25~a_0$ in the present work. The above
considerations give us a range of sensible values for $\alpha$ and
$\sigma$, but within this range physical properties still depend
on the chosen values. An approach is still needed to better resolve
these depending on the system. 

\section{CALIBRATION}

In principle, if the parameters $\alpha$ and $\sigma$ defining a
physical $V_{\rm e}$ were chosen correctly, an effective pressure
$P_{\rm eff}$ equal to the chosen input pressure $P_{\rm in}$ would be
felt within the nanocrystal. $V_{\rm e}$ could be determined for
different solvent-nanocrystal interfaces by comparison with
simulations using explicit solvents, e.g.\ with MD. This would however
not be practical in a fully \emph{ab initio } way as explained in
Section I. An empirical parametrization would also be difficult
considering the limited resolutions of experimental
methods. Alternatively, $\alpha$ and $\sigma$ can be calibrated by
comparing $P_{\rm eff}$ and $P_{\rm in}$ if a satisfactory definition
of $P_{\rm eff}$ is available.  This can be done by exploiting the
virial theorem in MD simulations\cite{1bealing2010constant}, but not
in geometry optimization calculations.  For these, a promising
approach is to exploit the experimental\cite{Hannah2012a} and
computational\cite{degoli2004ab} result that bonds in the bulk-like
core of sufficiently large alkane-terminated diamond phase Si
nanocrystals display elastic properties similar to the bulk for a
range of sizes. $P_{\rm eff}$ can then be estimated
for a range of systems and pressures from the
compression of core bonds after quasistatic geometry optimization.
\begin{figure*}[t]
\includegraphics[scale=0.34]{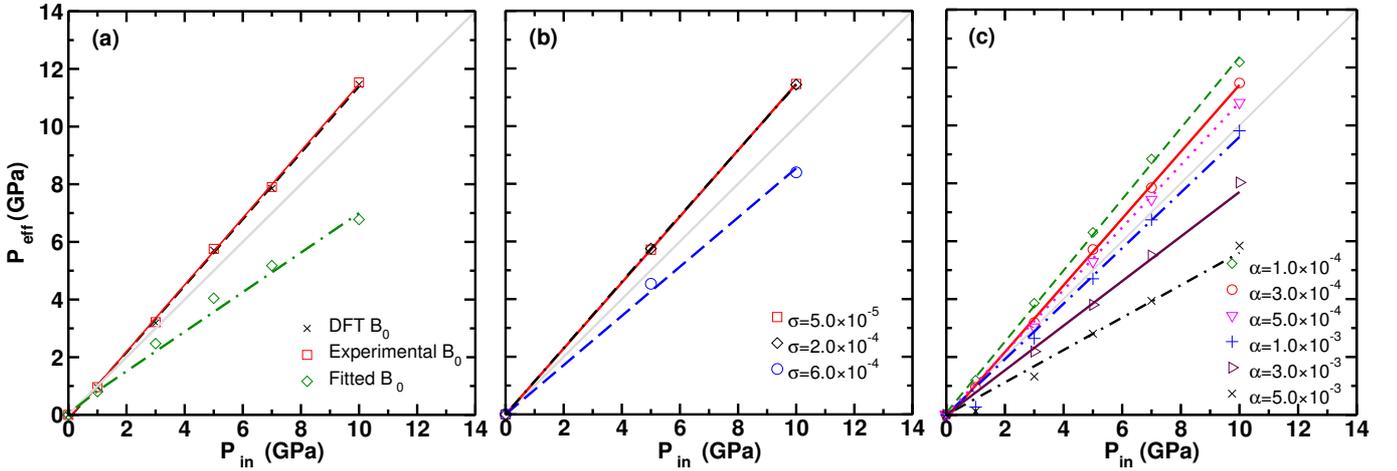}

\caption{\label{fig:Calibration}Calibration plots showing $P_{\rm
    eff}$ vs $P_{\rm in}$: (a) for different parametrization schemes
  of $B_0$, all for $\alpha=3\times10^{-4}~a_0^{-3}$ and
  $\sigma=5\times10^{-5}~a_0^{-3}$; (b) for different values of
  $\sigma$, all using the DFT bulk value of $B_0$ and
  $\alpha=3\times10^{-4}~a_0^{-3}$; (c) for different values of
  $\alpha$, all using the DFT bulk value of $B_0$ and
  $\sigma=5\times10^{-5}~a_0^{-3}$.}
\end{figure*}

Here we use the Vinet equation, with the bulk values of $B_0$ and $B_0'$
obtained as discussed in Section II, expressed in terms of the compressed
and equilibrium (0 GPa) bulk-like conventional lattice parameters
$a$ and $a_{\rm eq}$: 
\begin{align}
E(a) &= E_0+\frac{B_0 a_{\rm eq}^3}{(B_0'-1)^2}\Biggl(1-\frac{1}{2}\Biggl[\frac{3a}{a_{\rm eq}}\biggl(B_0'-1\biggl)-3B_0'+5\Biggl]\nonumber \\
& \times \exp\Biggl[-\frac{3}{2}(B_0'-1)\biggl(\frac{a}{a_{\rm eq}}-1\biggl)\Biggl]\Biggl).\label{eq:vinet press-1}
\end{align}
Defining $a$ for the nanocrystal 
in terms of averaged bond lengths for core
Si atoms (chosen as those atoms that are bonded exclusively to other
Si atoms) an effective pressure $P_{\rm eff}$ experienced by the nanocrystal
can be estimated from 
the volume derivative of the Vinet equation of state:
\begin{align}
P_{\rm eff}(a) &= 3B_0 \biggl(\frac{a_{\rm eq}}{a}\biggl)^2\biggl(1-\frac{a}{a_{\rm eq}}\biggl)\nonumber \\
& \times \exp\Biggl[-\frac{3}{2}(B_0'-1)\biggl(\frac{a}{a_{\rm eq}}-1\biggl)\Biggl].\label{eq:vinet press}
\end{align}
The Vinet equation holds in the absence of phase transitions and was
found to give similar, albeit better fitted, results than the Birch-Murnaghan
equation.\cite{birch1988elasticity} $P_{\rm eff}$ can then be
compared to the input pressure $P_{\rm in}$ that generated the
compression, thus allowing the calibration of $\alpha$ and $\sigma$. 

A hydrogenated tetrahedral ${\rm Si}_{71}{\rm H}_{60}$ nanocrystal in
the diamond phase was used for the calibration as it displays a
sizable core which behaves elastically like the bulk and justifies our
use of bulk values for $B_0$ and $B_0'$ in the calibration.  The
average Si--Si bond length is found to be contracted compared to the
bulk; the contraction is reduced and tends towards the bulk value as
the size of the nanocrystal is increased. Looking at individual Si--Si
bonds, it is found that the outer shell is contracted, which has been
interpreted as due to surface
stress,\cite{buttard1999porous,weissker2003structural} while inner
shells substantially agree with bulk values. Similar results have been
found for ${\rm Si}_{29}{\rm H}_{36}$ and ${\rm Si}_{35}{\rm H}_{36}$
using DFT with norm-conserving pseudopotentials and a local density
approximation exchange-correlation functional.\cite{degoli2004ab} In
classical linear elasticity, inhomogeneities, whether \emph{in vacuo}
or embedded in a material, have size-independent elastic
fields.\cite{eshelby1957determination} This is the result of
neglecting surface energies which can be justified when the ratio of
surface to volume atoms is small. At the nanoscale, however, this
ratio becomes important and the surface induces size-dependent elastic
fields that are long-range.\cite{sharma2004size} One would expect the
surfaces to induce a size-dependent strain-field and to distort the
core atoms for the nanocrystal sizes investigated in this
work. However, the stiffness of Si nanocrystals and the absence of
reconstruction of the hydrogen-passivated surfaces result in
distortions that are smaller than the displacement tolerance. This
limits the effect of the surfaces for the sizes considered and
simplifies the mapping between effective pressure and
compression. While the bond distributions changes with the selection
of the core atoms, the positions of peaks of the distribution are
found to be insensitive, within the displacement tolerance, to that
choice when excluding the outer shell atoms.

Figure~\ref{fig:Calibration} shows the results of our calibration
for a range of parameter choices. 
A procedure where $B_0$ and $B_0'$ were fitted separately
from Eq.~\ref{eq:vinet press-1} for each $\alpha$ (diamonds) is
seen to produce poor agreement between $P_{\rm eff}$ and $P_{\rm in}$.
By contrast, it was found that using fixed values of $B_0$ and $B_0'$,
either from DFT bulk values from CASTEP (crosses) or experimental values
(squares), produced very similar results, and the expected linear relationship
was observed (Fig.~\ref{fig:Calibration}(a)). 
This suggests that the assumptions entering our calibration
approach and the volume definition are valid.
Figure~\ref{fig:Calibration}(b) shows that for a fixed value of $\alpha$,
the $P_{\rm eff}$ curves converge as a function of $\sigma$, 
towards the $\sigma=5\times10^{-5}~a_0^{-3}$
line. This highlights the importance
of tuning $\sigma$ for an accurate definition of $V_{\rm e}$ as
discussed in Section II.
Finally, given an appropriate choice of $\sigma$, Fig.~\ref{fig:Calibration}(c)
shows that there is a dependence of the compression on the chosen $\alpha$.
This can be understood from the volume definition of Eq.~\ref{eq:Vdef}:
changing $\alpha$ corresponds
to using a different model of solvent-nanocrystal interaction, by altering
the electronic density up to which solvent molecules approach the
nanocrystal. The range of $\alpha$ values $3.0\times 10^{-4}$ to
$1.0\times 10^{-3}~a_0^{-3}$ is found to give agreement between
$P_{\rm eff}$ and $P_{\rm in}$ within 15\%. 

While the parameters were calibrated on the ${\rm Si}_{71}{\rm H}_{60}$
nanocrystal and model a representative solvent-nanocrystal interface, the parameters are
expected to be transferable to silicon nanocrystals of different sizes
and shapes, particularly if they have similar surface facets, surface
reconstructions or similar ligands. We illustrate this by comparing the
effective pressure $P_{\rm eff}$ obtained at a representative value of $P_{\rm in} = 10$~GPa,
with $\alpha=3.0\times 10^{-4}~a_0^{-3}$ and $\sigma=5.0\times 10^{-5}~a_0^{-3}$, for three different nanocrystal sizes:
${\rm Si}_{35}{\rm H}_{36}$, ${\rm Si}_{71}{\rm H}_{60}$ and ${\rm Si}_{181}{\rm H}_{110}$. We obtain $P_{\rm eff}$ = 11.9, 11.5 and 11.5~GPa respectively, indicating that the bulk-like cores of ${\rm Si}_{71}{\rm H}_{60}$ and ${\rm Si}_{181}{\rm H}_{110}$ have identical responses under pressure while ${\rm Si}_{35}{\rm H}_{36}$ displays a small discrepancy, which can be understood as due to the small nanocrystal size and consequent small region of bulk-like core. The transferability is expected to hold beyond hydrogenated silicon
because all semiconductor nanocrystal surfaces, with or without organic
ligands, will display a fairly similar range of values of valence electronic density
inside and a similar exponential decay rate outside the surface.\cite{engels1998} The
solvent-nanocrystal interface resulting from a given choice of
parameters can be considered appropriate so long as the lengthscale of
the isosurface variations is smaller than the size of the solvent molecules given by their van der Waals surface.

\section{RESULTS: PRESSURE-INDUCED TRANSFORMATIONS IN SILICON NANOCRYSTALS}

We now turn our attention to structural transformations in the Si
nanocrystals ${\rm Si}_{35}{\rm H}_{36}$ and ${\rm Si}_{71}{\rm H}_{60}$,
which have been studied by other methods.\cite{1martonak2000ab,1martovnak2001new,1molteni2001first,1Martonak2002,1molteni2005polyamorphism}
We then demonstrate capability for larger system sizes by studying
${\rm Si}_{181}{\rm H}_{110}$. Bulk Si is of great technological
importance and has been widely used as semiconductor in both its crystalline
and amorphous forms. Its phase diagram has been extensively studied
and, under pressure, bulk Si has been observed to transform between
12 different structures. It transforms from the cubic diamond to the
$\beta$-Sn phase at 11.7~GPa, followed by the Imma phase, primitive
hexagonal (ph), orthorhombic, hexagonal closed packed (hcp) and face
centered cubic (fcc) phases at respectively 13.2, 15.4, 38, 42 and
79~GPa.\cite{duclos1990experimental,katzke2007structural,olijnyk1984structural}
Upon pressure release the ph and $\beta$-Sn phases are observed, but
the diamond phase is not recovered upon full release of the pressure.
Instead, a host of crystalline and amorphous metastable structures
are observed, the most common of which are the BC8 and R8 phases that
correspond to distorted tetrahedral structures. Unlike the bulk, the
small size and the stabilising effect of the surfaces of Si nanocrystals
allows for metastable structures and transformation mechanisms that
are still the subject of investigation. Size-dependence of structural,
optical and electronic properties has been reported for a range of
nanocrystal sizes in both colloidal\cite{Tolbert1996a,Hannah2012a}
and porous forms\cite{deb2001pressure,garg2011memory}. 
\begin{figure*}[t]
\includegraphics[scale=0.5]{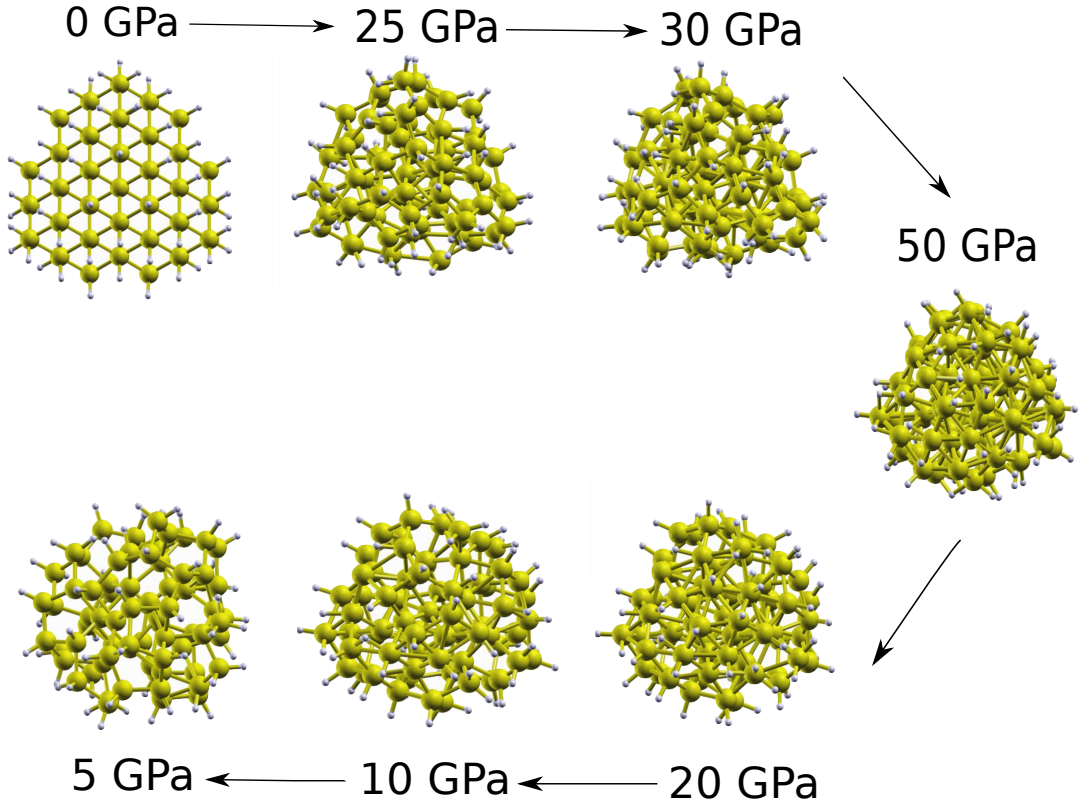}

\caption{\label{fig:Bond-distribution-and-1-1-1-1}Structures of ${\rm Si}_{71}{\rm H}_{60}$
at 0, 25, 30 and 50 GPa and on releasing pressure at 20, 10 and 5
GPa.}
\end{figure*}

Recent X-ray diffraction (XRD) experiments on colloidal plasma-synthesized
Si nanocrystals, where the surfaces are initially H terminated and
later functionalized with dodecane,  investigate nanocrystals of diameters 3.2, 3.8
and 4.5~nm,
under pressures in the range 0--73~GPa
at room temperature.\cite{Hannah2012a} A transformation between the diamond
and what is interpreted to be the ph structure is reported in the
range 17--22~GPa although the small size of the sample results in
significant broadening of the spectra and makes it difficult to identify
the structure. Another structural transformation occurs in the range
40--44~GPa and matches an hcp structure. The ph structure is recovered
upon decompression down to as low as 18.4~GPa followed by a stable
amorphous structure upon complete decompression. 

XRD and Raman spectroscopy experiments\cite{deb2001pressure} performed
on porous Si with average crystallite diameters $\sim$5~nm (with distributions at 3~nm and 7~nm)
report a transformation from diamond to a high-density amorphous (HDA) phase at 14~GPa which, upon pressure release, transforms to a low-density amorphous (LDA). More recent work \cite{garg2011memory} on porous Si with crystallites of $\sim$4~nm average diameter observe a transformation from  diamond to  ph phase at 20~GPa with no
amorphisation up to 39~GPa. A HDA phase is recovered upon decompression around 15~GPa followed by an LDA phase at 4.5~GPa. Under a further pressurization
cycle, an amorphous to crystalline transformation is observed between
LDA and ph at 18~GPa. Such reversible transformations between LDA,
HDA and ph phases have also been observed for amorphous bulk Si
(a-Si).\cite{pandey2011pressure,daisenberger2007high,daisenberger2011polyamorphic} 

Theoretical investigations have attempted to characterize these amorphous
phases in bulk Si\cite{morishita2004high,morishita2009structural,durandurdu2006ab,durandurdu2001ab,daisenberger2007high,daisenberger2011polyamorphic}
and hydrogenated Si nanocrystals.\cite{1martonak2000ab,1martovnak2001new,1molteni2001first,1Martonak2002,1molteni2005polyamorphism}
The LDA phase has been described as a disordered tetrahedrally coordinated
network, the HDA as a deformed tetrahedral network\cite{morishita2004high}
with the presence of interstitials increasing the coordination to
5--6 and finally a very-high-density amorphous phase (VHDA) has been
postulated\cite{durandurdu2001ab} with coordinations 8--9 as found
also in ice.\cite{1martovnak2004polyamorphism} 
This classification of the amorphous structures is adopted in this paper.

Structural properties are generally analysed in terms of bond-length
distribution, coordination number and bond-angle distribution. Moreover,
we employ
ring statistics as a way of tracking the evolution of the covalent
Si networks. Every Si atom can be treated as a node and
bonds as links connecting the nodes. We define an $n$-membered ring
as a closed path connecting $n$ atoms according to Guttman's definition,
focusing on the total number of rings $R_{\rm T}$ and the proportion
of nodes belonging to at least one $n$-membered ring $P_n$.\cite{guttman1990ring,le2010ring}
Two atoms are considered to be bonded when separated
by a distance smaller than the first minimum of the radial
distribution function of the bulk $R_{{\rm c}}=5.33~a_0$.
All nanocrystals were initially relaxed with geometry optimization
at 0~GPa, and then pressure was applied in steps of 5~GPa or less,
relaxing the geometry at every pressure to find the minimal enthalpy
configuration. ${\rm Si}_{35}{\rm H}_{36}$ and ${\rm Si}_{71}{\rm H}_{60}$
were investigated in a pressure range 0--50~GPa while only 0--25~GPa
was considered for ${\rm Si}_{181}{\rm H}_{110}$ as the system becomes
metallic beyond 25~GPa and occupancy smearing would be needed. 

A density cutoff value of $\alpha=3.0\times10^{-4}~a_0^{-3}$
was chosen for the rest of the calculations because shown to result
in a good calibration (Fig.~\ref{fig:Calibration}) and as enables
direct comparison with the simulations of Cococcioni \emph{et~al.}
\cite{Cococcioni2005}
\begin{figure*}
\includegraphics[scale=0.3]{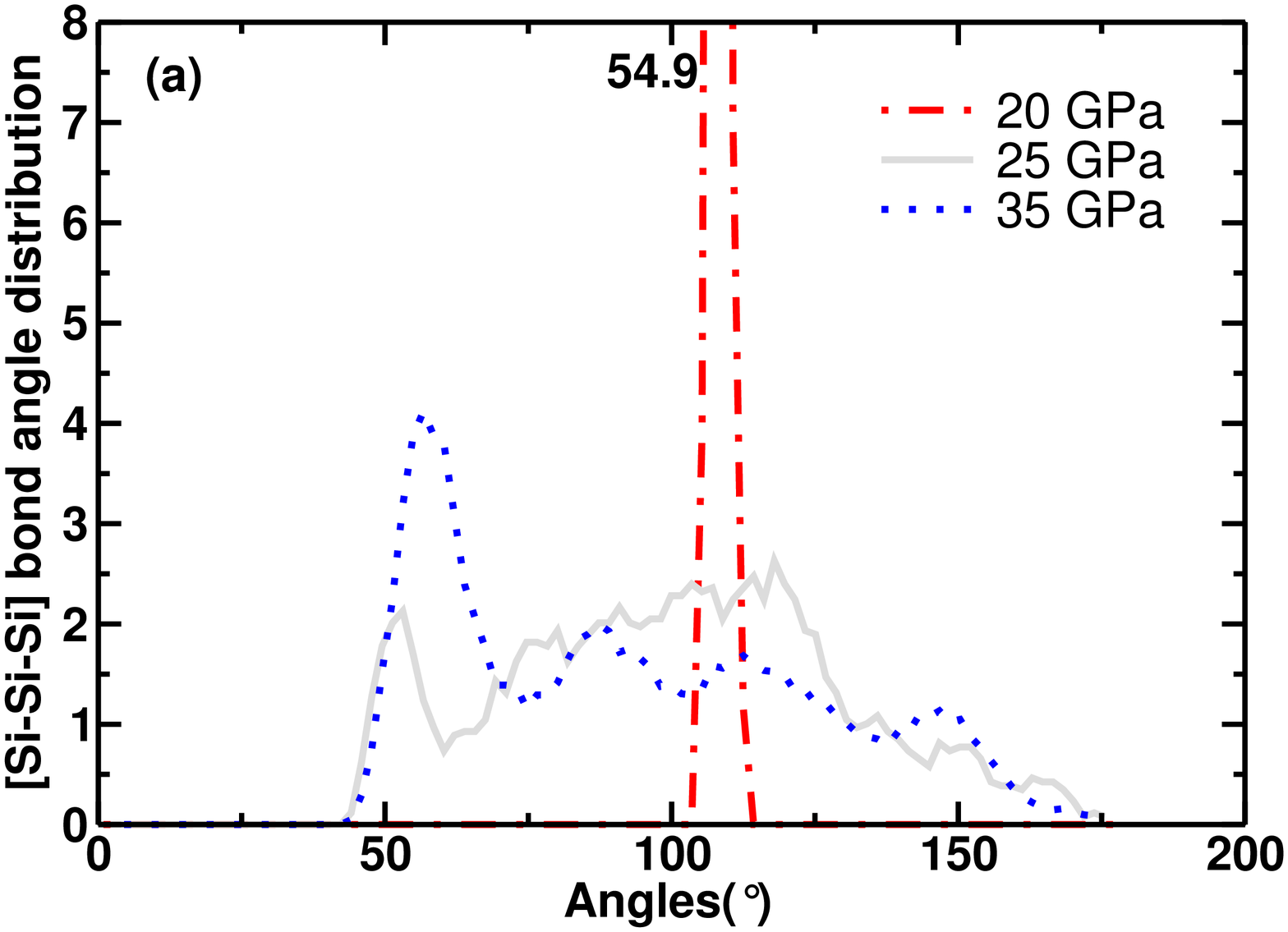}\includegraphics[scale=0.3]{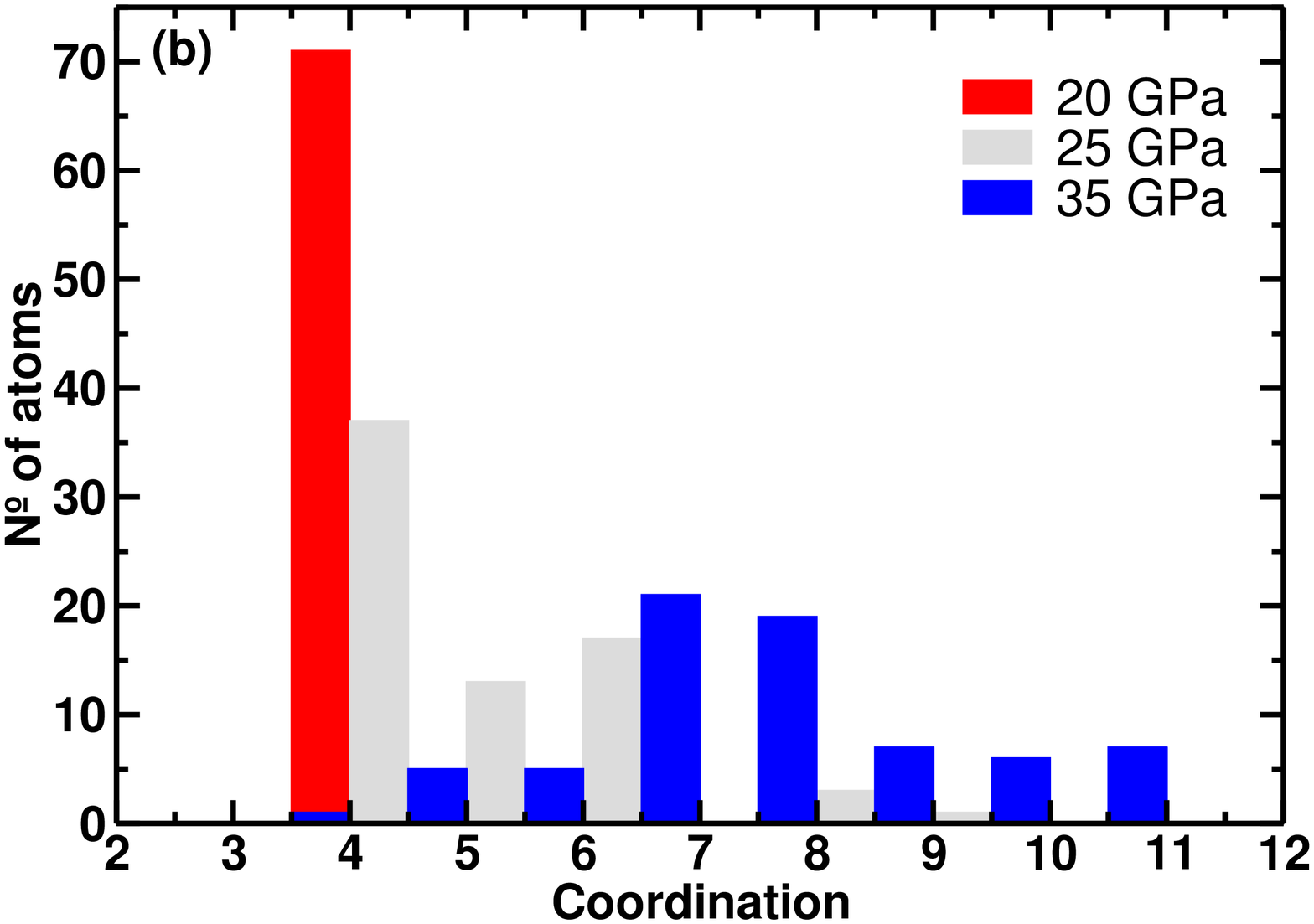}

\includegraphics[scale=0.3]{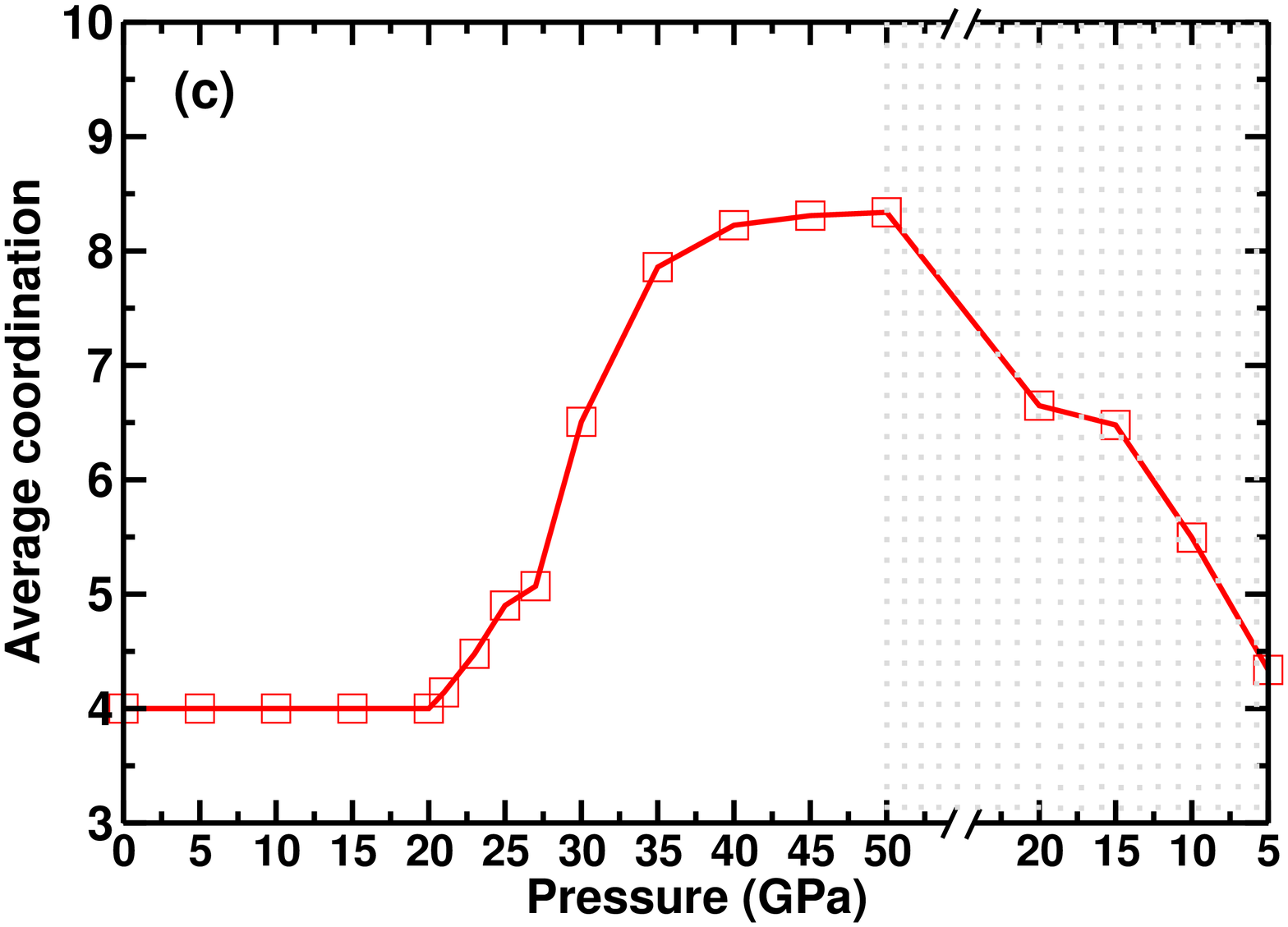}\includegraphics[scale=0.3]{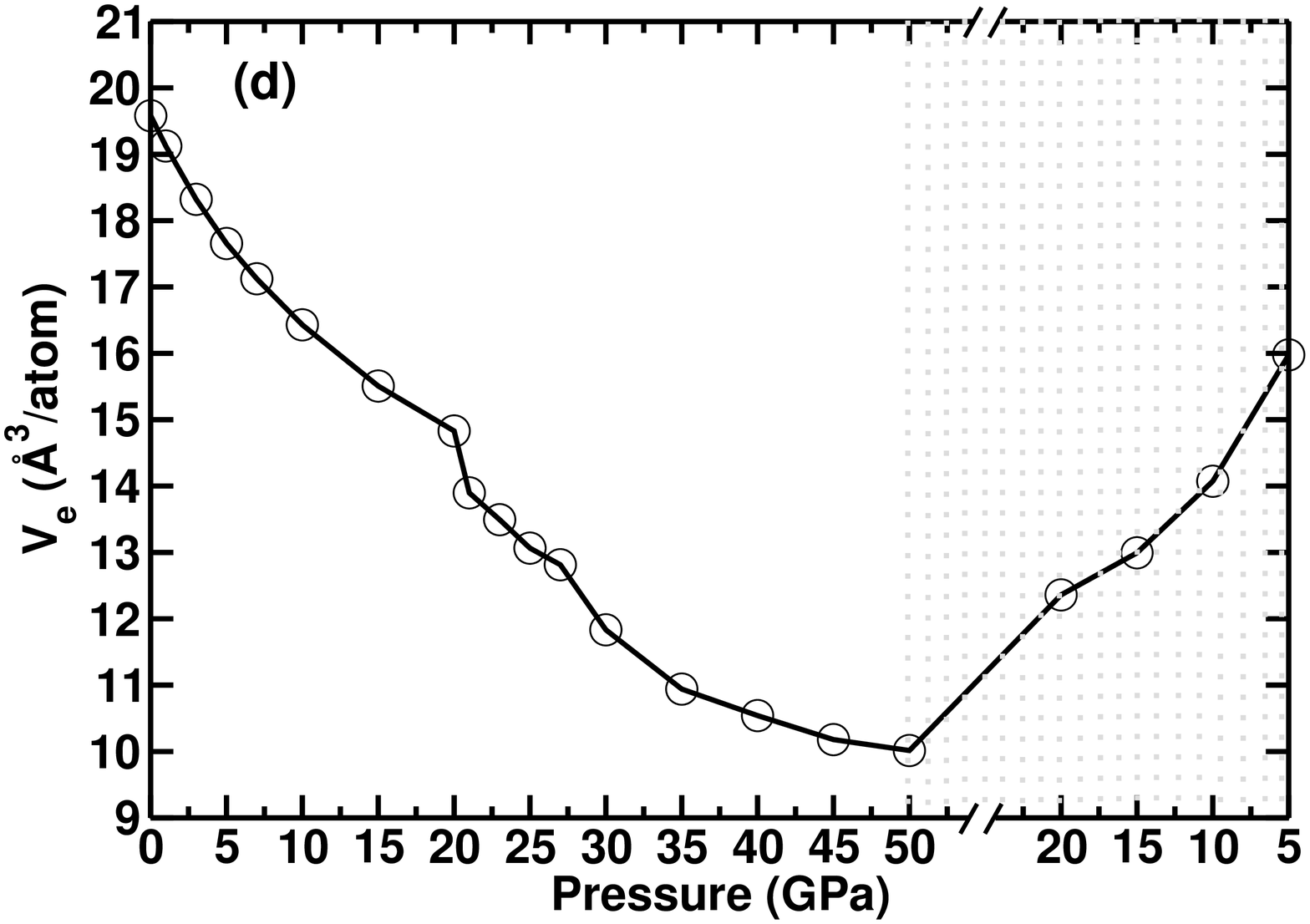}

\includegraphics[scale=0.3]{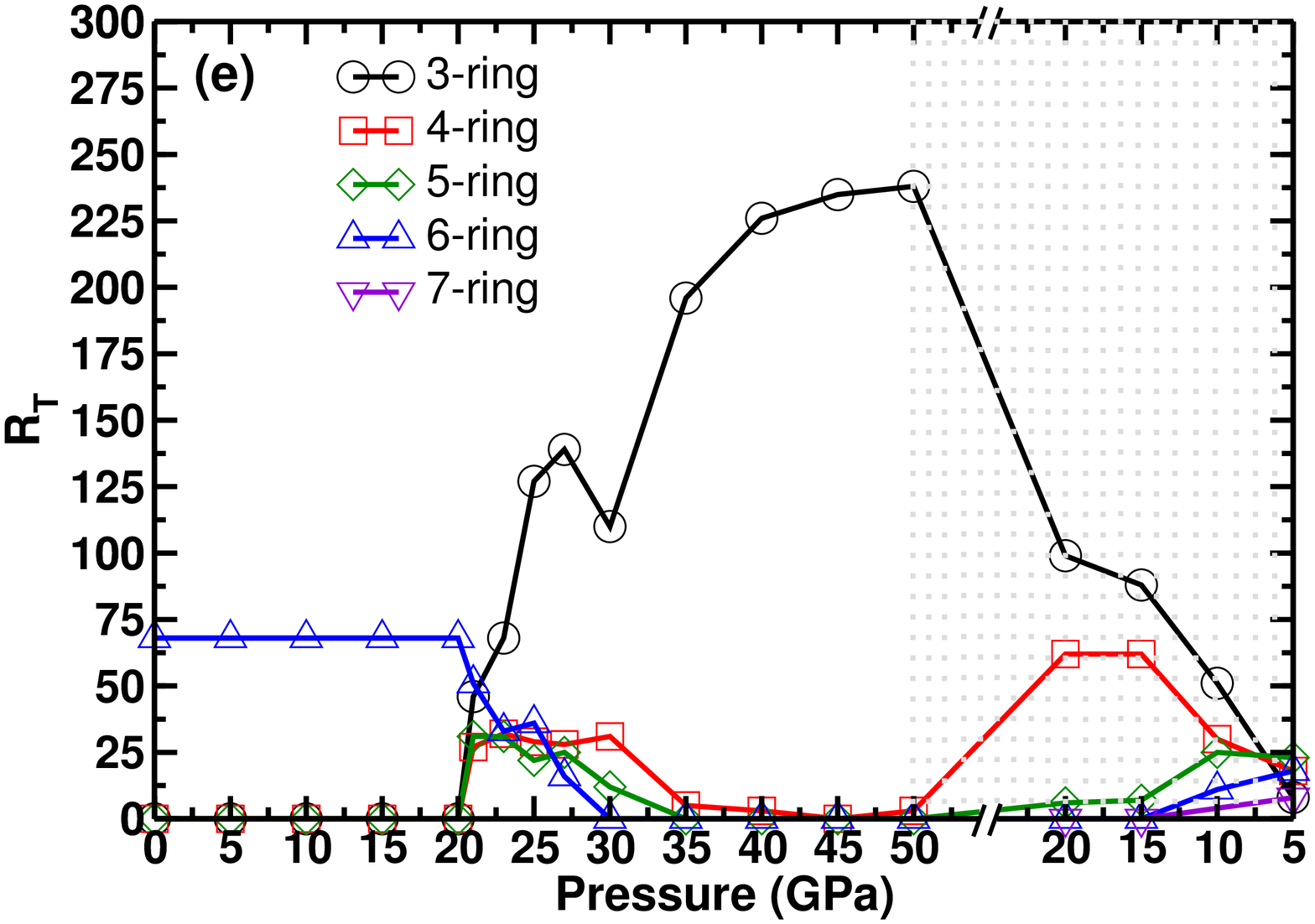}\includegraphics[scale=0.3]{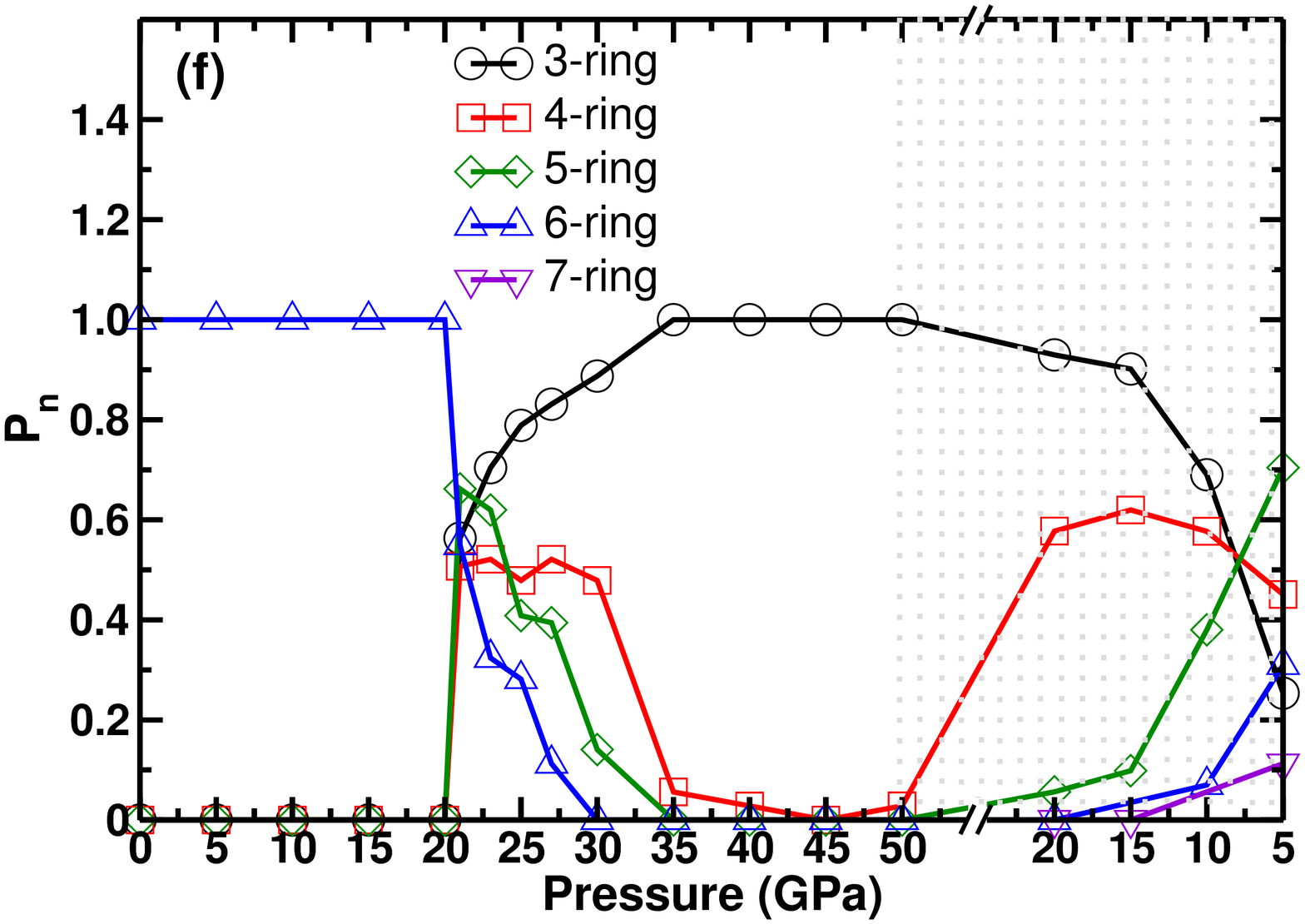}

\caption{\label{fig:Structural info} Structural transformations of ${\rm Si}_{71}{\rm H}_{60}$
under pressure (the shaded region corresponds to the depressurization):
(a) distribution of the bonded Si--Si--Si angles at 20, 25 and 35~GPa;
(b) distribution of the coordination numbers of Si atoms at 20, 25
and 35~GPa; (c) total number of rings $R_{\rm T}$ with pressure; (d)
proportion of nodes belonging to at least one $n$-membered ring $P_n$;
(e) average coordination number of Si atoms with pressure; (f) electronic
volume per atom with pressure. }
\end{figure*}
 
Figure~\ref{fig:Bond-distribution-and-1-1-1-1} shows the structures
of ${\rm Si}_{71}{\rm H}_{60}$ as it is compressed in the range 0--50~GPa
and depressurized to 5~GPa, while Fig.~\ref{fig:Structural info} shows a
range of descriptors of these transitions. The bond-angle distribution
in Fig.~\ref{fig:Structural info}(a) initially shows a single peak at
$109.5^{\circ}$, typical of the tetrahedral diamond structure. The peak
broadens when the pressure applied increases from 0~GPa to 20~GPa, which reflects
a lower degree of symmetry in the structure. No change in the coordination
number of Si atoms is observed at this stage and a tetrahedral coordination
is retained for the innermost Si atoms. When the pressure applied
is increased to $\sim$35~GPa, a structural change is observed as
the nanocrystal amorphizes, as evidenced by the broad bond-angle distribution.
The transformation is mediated by the breaking and making of bonds
which are accompanied by the appearance of interstitial atoms and
an increase in coordination numbers (Fig.~\ref{fig:Structural info}(b)).
To better resolve the transformation, calculations for ${\rm Si}_{71}{\rm H}_{60}$
were repeated in steps of 2~GPa in the interval 20--30~GPa. Between
21--30~GPa, the average coordination of Si atoms (Fig.~\ref{fig:Structural info}(c))
increases from 4 at 0--20~GPa, to 5.1 at 27~GPa which is consistent
with a transformation from diamond to HDA. However, even at 30~GPa
the nearest neighbour peak remains, suggesting that some local order
is retained and the first coordination shell is preserved. As the
pressure is increased further, the average coordination reaches 8.3
at 50~GPa which matches that of a VHDA structure. Upon pressure release,
the average coordination plummets to 6.7 at 20~GPa and 4.3 at 5~GPa
corresponding to a disordered tetrahedral network consistent with
an LDA structure. Similarly, for ${\rm Si}_{35}{\rm H}_{36}$ one
obtains a coordination of 4 at 0~GPa, 5.5 at 30~GPa, 7.3 at 40~GPa,
7.7 at 50~GPa, 4.5 upon decompression to 5~GPa and for ${\rm Si}_{181}{\rm H}_{110}$
4 at 0~GPa and 5.5 at 25~GPa. Fig.~\ref{fig:Structural info}(d) shows
the electronic volume per atom for ${\rm Si}_{71}{\rm H}_{60}$.
This reveals discontinuous changes in the intervals 20--21~GPa and again
at 27--30~GPa, which suggests that the structural transformations from
diamond to HDA and HDA to VHDA are first order. 

Our results are consistent with previously-reported Car-Parrinello MD
simulations on ${\rm Si}_{35}{\rm H}_{36}$ and ${\rm Si}_{71}{\rm H}_{60}$ at
600~K using a classical explicit soft-sphere solvent\cite{1molteni2001first}
and on ${\rm Si}_{35}{\rm H}_{36}$ at 300~K using the electronic enthalpy method\cite{Cococcioni2005}. In the former simulations,
transformations from the diamond to an amorphous structure with average
coordination of core Si atoms of 7.3 were reported
at about 30~GPa for ${\rm Si}_{71}{\rm H}_{60}$ and 35~GPa for ${\rm Si}_{35}{\rm H}_{36}$.
An amorphous structure with average coordination 4.3 is recovered
upon decompression to 5~GPa. In the latter simulations, 
${\rm Si}_{35}{\rm H}_{36}$
is found to amorphize around 26--28~GPa with average coordination
of Si atoms reaching $\sim$6.5 and to remain amorphous upon pressure
release to 0~GPa with an average coordination of $\sim$4. No sensitivity
of the results was reported when changing $\alpha$. This can be understood
as resulting from the small range of values chosen, combined with thermal
noise concealing the dependence on parameters seen in the present
work. Differences in transformation pressure for the same system
between MD simulations using 
an explicit solvent\cite{1molteni2001first} 
and the electronic enthalpy method\cite{Cococcioni2005} are due to the
different way that pressure is applied.\cite{1bealing2010constant}
We do not expect direct agreement in the transformation pressure
with MD nor experimental results
considering that we have used a quasistatic approach: in the absence
of thermal fluctuations, one needs to overpressurize the system to
overcome the large activation energies associated with the energetic
cost of making and breaking bonds. The absence of thermal noise in
our simulations does, however, facilitate the detailed monitoring
of the amorphization
and provides complementary information to that
obtained with MD.

The ring statistics shown in Fig.~\ref{fig:Structural info}(e) and 
Fig.~\ref{fig:Structural info}(f)
indicate the presence of four distinct regions. In the interval
0--20~GPa, only 6-membered rings are present. The population of 6-membered
rings decays in favour of 3-, 4- and 5-membered rings in the interval
21--30~GPa. Between 30--50~GPa, the 3-membered ring population grows
further at the expense of 4- and 5-membered rings and dominates at
50~GPa. Upon pressure release, the 4-, 5-, 6- and 7-membered ring
population recovers while the 3-membered ring population drops sharply
as the nanocrystal dilates. 6-membered rings are a signature of the
corner-sharing tetrahedra in the diamond cubic structure, while 3-membered
rings arise through the formation of the equilateral triangles that cause
the peak at $60^{\circ}$ in the bond-angle distribution (Fig.~\ref{fig:Structural info}(a)).
The presence of 3-, 4- and 5-membered rings indicates amorphization
and the larger 7-membered ring the formation of voids (Fig.~\ref{fig:-3--(yellow),}).
\begin{figure}
\includegraphics[scale=0.425]{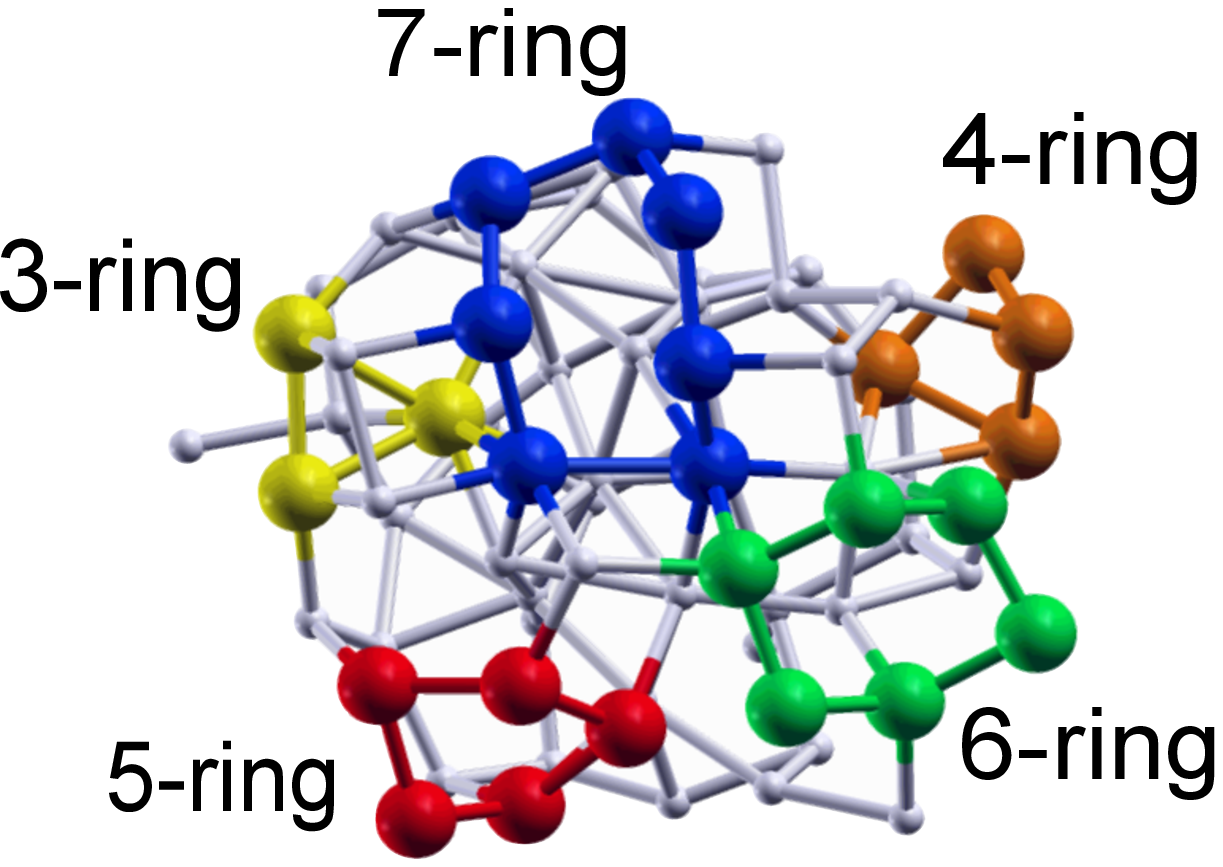}\caption{\label{fig:-3--(yellow),} 3- (yellow), 4- (orange), 5- (red), 6-
(green) and 7- (blue) membered rings in ${\rm Si}_{71}{\rm H}_{60}$
upon pressure release to 5~GPa.}
\end{figure}
Our results are consistent with the existence of three amorphous
structures visited during the pressure-induced structural transformation: 
HDA corresponding to average coordination numbers $\sim$5--6,
VHDA with coordinations $\sim$8--9 and LDA with coordination $\sim$4
obtained upon pressure release.

\begin{figure}
\includegraphics[scale=0.35]{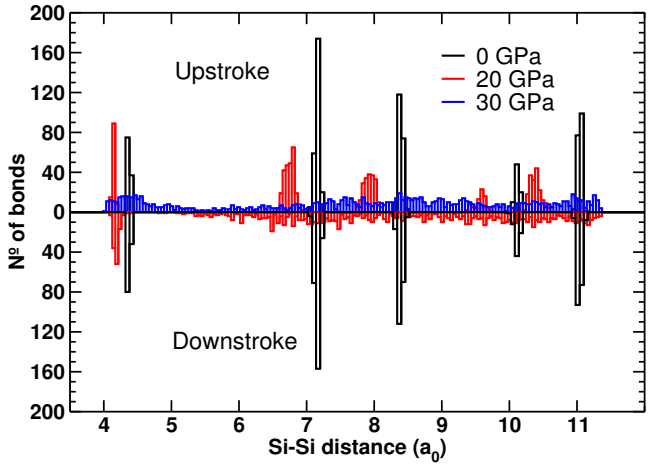}\caption{\label{fig:Bond-length-distribution}Distribution of Si--Si distances
in ${\rm Si}_{71}{\rm H}_{60}$ under a pressure cycle at pressures
0, 20 and 30 GPa upstroke (top panel) and downstroke starting from
30 GPa (bottom panel).}

\end{figure}

A reversible amorphization diamond$\rightarrow$HDA$\rightarrow$diamond
is obtained when performing a pressure cycle between 0 and 30~GPa
(Fig.~\ref{fig:Bond-length-distribution}). By contrast, when increasing
the pressure all the way to 50~GPa, irreversible bonding events accompany
the transformation, which proceeds as diamond$\rightarrow$VHDA$\rightarrow$LDA.
The final LDA structure is found to be higher in energy compared to the
original diamond structure and thus corresponds to a local minimum in energy. The
nanocrystals display hysteresis and comparing the upstroke and downstroke
20~GPa structures, it is clear that they are respectively crystalline
and amorphous.
This behavior demonstrates the possibility of trapping nanocrystals
in unconventional bonding geometries when performing a pressure cycle,
yielding electronic properties different from the bulk. Of great promise
is the possibility of designing materials with desired properties
using pressure as a way to tune these. 

\begin{figure}
\includegraphics[scale=0.35]{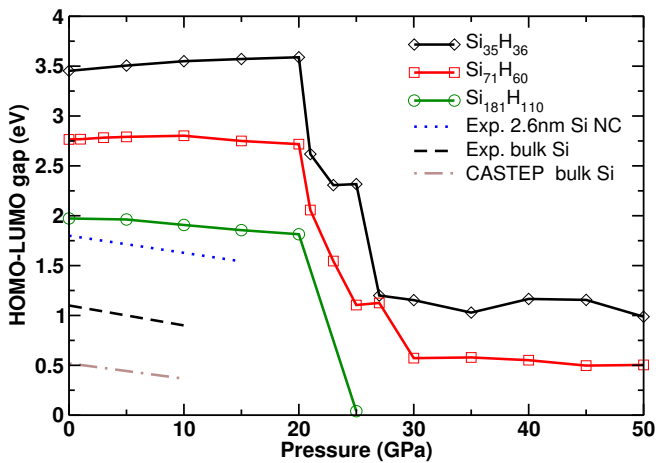}\caption{\label{fig:Trends-of-the-1-1-1}HOMO--LUMO gaps of ${\rm Si}_{35}{\rm H}_{36}$,
${\rm Si}_{71}{\rm H}_{60}$ and ${\rm Si}_{181}{\rm H}_{110}$ under
pressure and comparison with experiment \cite{Hannah2012a,welber1975dependence}
and 
DFT calculations of bulk Si in the diamond structure. }
\end{figure}

In particular, it is found that the HOMO--LUMO gap changes dramatically
under pressure and the way this happens is strongly size-dependent.
While it is well known that the local density approximation underestimates
the size of the gap for Si, it does give qualitative information
and significant trends. From Fig.~\ref{fig:Trends-of-the-1-1-1}
we observe that beyond a certain pressure, the gap drops sharply to
a lower value. All clusters are semiconductors at low pressure with
increased gaps for the smaller nanocrystals compared to the bulk.
This can be understood as due to quantum confinement which is significant
for the nanocrystals under investigation. The ${\rm Si}_{181}{\rm H}_{110}$
nanocrystal gap sharply drops to 0.04~eV at $\sim$25~GPa showing
that it becomes metallic, whereas the smaller clusters retain a sizeable
gap even at higher pressures. In the pressure range 0--20~GPa, the
gap of ${\rm Si}_{35}{\rm H}_{36}$ increases with a slope that reduces
as the pressure is ramped up. Meanwhile, for ${\rm Si}_{71}{\rm H}_{60}$,
the gap increases at first and decreases in the range 10--20~GPa.
For ${\rm Si}_{181}{\rm H}_{110}$, the gap decreases linearly in
the range 5--20~GPa. This size-dependent behaviour can be interpreted
as a competition between quantum confinement and the negative pressure
coefficient of diamond Si. The former tends to increase the gap
as the nanocrystals are compressed, while the latter tends to decrease
it due to the dominance of the indirect transition (corresponding
to ${\rm X}_{{\rm conduction}}-\Gamma_{{\rm valence}}$ in the bulk). ${\rm Si}_{181}{\rm H}_{110}$,
of diameter 2.2~nm, has a change of gap with pressure of -10.7~meV~GPa$^{-1}$
which is of the same order as the experimental results for 2.6~nm
diameter nanocrystals of -17.2~meV~GPa$^{-1}$.\cite{Hannah2012a} As the
nanocrystal size is increased, the quantum confinement effect is expected
to vanish and a linear decrease of the gap remain with a slope tending
to the 
DFT value for bulk diamond Si of -15.4~meV~GPa$^{-1}$. The above
results highlight the capability of our approach to simulate sizes
of experimental relevance with 
DFT accuracy.

\section{CONCLUSION}

We have implemented an electronic enthalpy method in a linear-scaling
DFT code (ONETEP) to simulate nanocrystals under pressure. An approach to calibrate
the parameters defining the electronic volume in the context of geometry
optimizations was proposed, demonstrating how the pressure felt inside
the nanocrystal can be successfully matched to the input pressure
in the electronic enthalpy functional. We have applied this method
to the structural transformations of hydrogenated Si nanocrystals
of different sizes and obtained results in good agreement with simulations
using explicit solvents. Our quasistatic investigation has allowed
for the detailed study of polyamorphic transformations and provided
information that would be difficult to extract with the thermal noise
of a MD simulation. Size-dependent structural transformations were
obtained between the diamond structure and the amorphous LDA, HDA
and VHDA structures. The behaviour of the intermediate structures
upon pressure release was investigated and depressurizing from HDA
and VHDA structures was found to give respectively diamond and LDA
structures. These have distinct electronic properties and the changes
in HOMO--LUMO gaps with pressure were analyzed for different nanocrystal
sizes and display qualitative agreement with experiment of similar
diameters. The present work highlights the capability of our approach; barring
further progress in the synthesis and probing of smaller nanocrystal
sizes, techniques such as linear-scaling DFT become essential to simulate
sizes of experimental relevance with quantum accuracy.
\begin{acknowledgments}
The authors would like to thank Francesco Mauri (Universit\'e Pierre et Marie Curie, Paris, France) 
for a helpful discussion.
The authors are grateful for the computing resources provided by the Imperial
College High Performance Computing Service and the HECToR UK
supercomputer facility which have enabled all the simulations presented
here. The HECToR computer time was funded by  the UK Engineering and Physical Sciences Research Council (EPSRC) grant EP/F037457/1 and a RAP class 1b award. N.R.C.C. was supported through a studentship in the Centre for
Doctoral Training on Theory and Simulation of Materials at Imperial
College funded by EPSRC grant number EP/G036888/1. N.D.M.H. acknowledges the support
of EPSRC grant EP/G05567X/1, a Leverhulme early career fellowship
and the Winton Programme for the Physics of Sustainability. P.D.H. acknowledges
the support of a Royal Society University Research Fellowship. 
\end{acknowledgments}

\end{document}